\documentclass[12pt]{article}
\pdfoutput=1
\usepackage{amsmath,amssymb,bm,graphicx,url}

\begin{document}

\begin{titlepage}

\begin{flushright}
WSU-HEP-2001\\
April 6, 2020\\
\end{flushright}

\vspace{0.7cm}
\begin{center}
\Large\bf\boldmath
Model Independent Extraction of the Proton Charge Radius from PRad data 
\unboldmath
\end{center}

\vspace{0.8cm}
\begin{center}
{\sc Gil Paz}\\
\vspace{0.4cm}
{\it 
Department of Physics and Astronomy \\
Wayne State University, Detroit, Michigan 48201, USA 
}
\end{center}

\vspace{1.0cm}
\begin{abstract}
  \vspace{0.2cm}
  \noindent
The proton radius puzzle has motivated several new experiments that aim to extract the proton charge radius and resolve the puzzle. Recently PRad, a new electron-proton scattering experiment at Jefferson Lab, reported a proton charge radius of  $0.831\pm 0.007_\textnormal{statistical}\pm 0.012_\textnormal{systematic}$. The value was obtained by using a rational function model for the proton electric form factor. We perform a model-independent extraction using $z$-expansion of the proton charge radius from PRad data. We find that the model-independent statistical error is more than 50\% larger compared to the statistical error reported by PRad.

\end{abstract}
\vfil

\end{titlepage}

\section{Introduction}
The proton is a composite particle. One way to define its size is by the proton charge radius, $r_E^p$. It is related to the slope of the proton electric form factor, $G_E^p$, at $q^2=0$, see (\ref{rpDefined}) below.  Since $G_E^p$ is a non-perturbative function of $q^2$, its slope must be extracted from data.  The most direct way to measure $r_E^p$ is by extracting $G_E^p$ from lepton-proton scattering and finding its slope at $q^2=0$. An indirect way is by using atomic spectroscopy.

Thus we have four different methods to extract $r_E^p$ from data: $e-p$ scattering, $\mu-p$ scattering, $e-p$ spectroscopy, and  $\mu-p$ spectroscopy. A fifth method, Lattice QCD, should become competitive in the future, see, e.g., \cite{Alexandrou:2020aja}. While $e-p$ scattering and spectroscopy extractions were available for a long time, $\mu-p$ spectroscopy only became available in 2010 from the work of the CREMA collaboration \cite{Pohl:2010zza, Antognini:1900ns}. Results from $\mu-p$ scattering are expected in the near future from the MUSE collaboration \cite{Gilman:2017hdr}. Ideally, all methods should give consistent results.  Surprisingly, in 2010, $\mu-p$ spectroscopy gave a value, $0.84184(67)$ fm,  that was considerably smaller than the CODATA value, $0.8768(69)$ fm \cite{Mohr:2008fa}. This difference is referred to as the ``proton radius puzzle". For a recent review, see \cite{Paz:2019wfq}. 

The puzzle has motivated  new theoretical and experimental work. Three new $e-p$ spectroscopy measurements were published recently.  
Two agree with the smaller value \cite{Beyer:2017,Bezginov:2019}, and one \cite{Fleurbaey:2018fih} with the larger value.  Two new $e-p$ scattering experiments, ISR and PRad, have published their results and more experiments are planned \cite{PRP2018}. ISR found $0.81(8)$ fm \cite{Mihovilovic:2016rkr}, and more recently \cite{Mihovilovic:2019jiz} $0.87(4)$ fm, that cannot distinguish between the two values. PRad  found  \cite{Xiong:2019umf} $0.831\pm 0.007_\textnormal{statistical}\pm 0.012_\textnormal{systematic}$ fm, which favors the smaller value. 

A main issue in extracting $r_E^p$ from scattering data is the unknown functional form of $G_E^p$. Recent extractions have used: dipole \cite{Higinbotham:2015rja}, polynomial \cite{Griffioen:2015hta, Horbatsch:2016ilr}, continued fraction \cite{Griffioen:2015hta}, modified $z$ expansion \cite{Horbatsch:2015qda},  or more complicated forms \cite{Alarcon:2018zbz}. For pre-2010 extractions see  \cite{Nakamura:2010zzi}. Different assumed functional forms can lead to \emph{different} radii and uncertainties from the \emph{same} data.  An alternative approach is the so-called $z$ expansion that only uses the known analytic structure of $G_E^p$. The $z$ expansion is the default method for meson form factors. It was first applied to baryon form factors in \cite{Hill:2010yb}. Extractions of $r_E^p$ using $z$ expansion favor the larger value \cite{Hill:2010yb,Lee:2015jqa}.

The default functional form for $G_E^p$ used by PRad is a rational function called  ``Rational (1,1)", see (\ref{Rational11}) below. Apart from the overall normalization (that does not affect the slope) it depends on two parameters. In \cite{Hill:2010yb} it was shown that a fit with a small number of parameters can underestimate the errors.  In figure S15 of the supplementary material of the PRad paper \cite{Xiong:2019umf}, the ``Rational (1,1)" fit and ``$2^\textnormal{ nd}$ order $z$-tran." give similar radii with similar uncertainty, but  ``$3^\textnormal{ rd}$ order $z$-tran." has twice the uncertainty. As was shown in \cite{Hill:2010yb}, adding higher powers of  $z$ without bounding the coefficients will cause the uncertainty to grow without bound. On the other hand, if we bound the coefficients, we obtain an extraction of $r_E^p$ that is independent of the number of the parameters we fit \cite{Hill:2010yb}.  Since the form factor must have the correct analytic structure and therefore can be expanded as a power series in $z$, we obtain an extraction of $r_E^p$ that is independent of the exact unknown functional form of the form factor. 

The goal of this paper is to perform such a model-independent analysis to the published PRad data and to see how it affects the errors on the extracted $r_E^p$. For simplicity, we use the values of $G_E^p$ reported by PRad in \cite{PRadData} and use only the  statistical errors\footnote{Determining the systematic error for the charge radius is much more involved and described in the supplementary material of the PRad paper \cite{Xiong:2019umf}.}. The rest of the paper is organized as follows. In section \ref{sec:FF} we briefly review the relevant form factor parameterization and the $z$ expansion. In section \ref{sec:PRad} we repeat the fits performed by PRad to its data and reproduce their results. In section \ref{sec:zfit} we perform a model-independent $z$-expansion fit to the PRad data and extract $r_E^p$. We present our conclusions in section \ref{sec:conclusions}.

\section{Form factor parameterization and the $\bm z$ expansion} \label{sec:FF}
The one-photon probe of the proton gives rise to two form factors, $F^p_1$ and $F^p_2$, 
\begin{equation}
\langle P(p')|J_\mu^{\rm em}|P(p)\rangle=\bar u(p')
\left[\gamma_\mu F^p_1(q^2)+\frac{i\sigma_{\mu\nu}}{2m_p}F^p_2(q^2)q^\nu\right]u(p)\,,
\end{equation} 
where $q^2=(p'-p)^2\equiv t\equiv -Q^2$. The electric and magnetic form factors are defined as \cite{Ernst:1960zza} $G^p_E(q^2) = F^p_1(q^2) + q^2 F^p_2(q^2)/4m_p^2$ and  $G^p_M(q^2) = F^p_1(q^2) + F^p_2(q^2)$. The proton charge radius squared is defined via the slope of $G_E^p(q^2)$ at $q^2=0$:
\begin{equation}\label{rpDefined}
\langle r^2 \rangle_E^p=
\frac{6}{G_E^p(0)}\frac{d}{dq^2}G_E^p(q^2)\bigg|_{q^2=0}\, .
\end{equation}
The proton charge radius is given by $r_E^p\equiv\sqrt{\langle r^2 \rangle_E^p}$. In \cite{Xiong:2019umf} $r_E^p$ is denoted by  $r_p$. 

$G_E^p(q^2)$ is analytic in the complex $q^2$ plane outside a cut that starts at the two-pion threshold $q^2=4m_\pi^2$.  The domain of analyticity can be mapped onto the unit $|z|<1$ circle via the transformation 
\begin{equation}\label{zdefind}
z(t,t_\textnormal{cut},t_0) = {\sqrt{t_\textnormal{cut}- t} - \sqrt{t_\textnormal{cut} - t_0} \over \sqrt{t_\textnormal{cut} - t} + \sqrt{t_\textnormal{cut} - t_0}  }\,,
\end{equation}
where $t_\textnormal{cut}=4m_\pi^2$ and $t_0$ determines the location of $z=0$. In the following we use $t_0=0$. In the $|z|<1$ unit circle $G^p_{E}$ is analytic and can be expanded as a power series:
\begin{equation}\label{zseries}
G^p_{E}(q^2)=\sum_{k=0}^\infty a_k\, z(q^2)^k.
\end{equation} 
The choice $t_0=0$ implies that $r_E^p$ depends only on $a_1$.

\begin{figure}
\begin{center}
\includegraphics[height=11.5em]{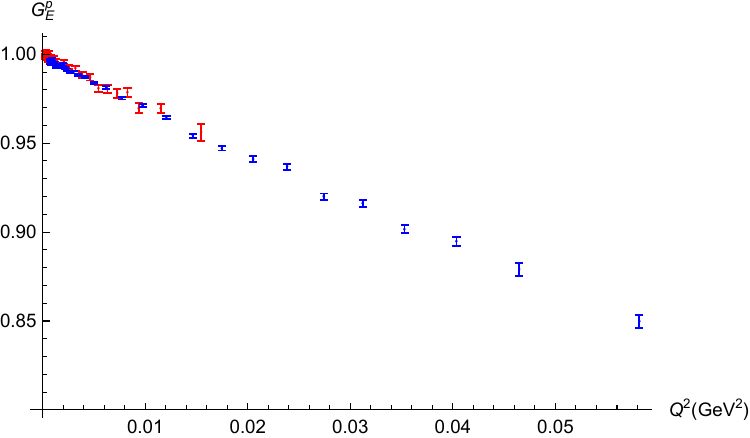}
\hspace{1cm}
\includegraphics[height=11.5em]{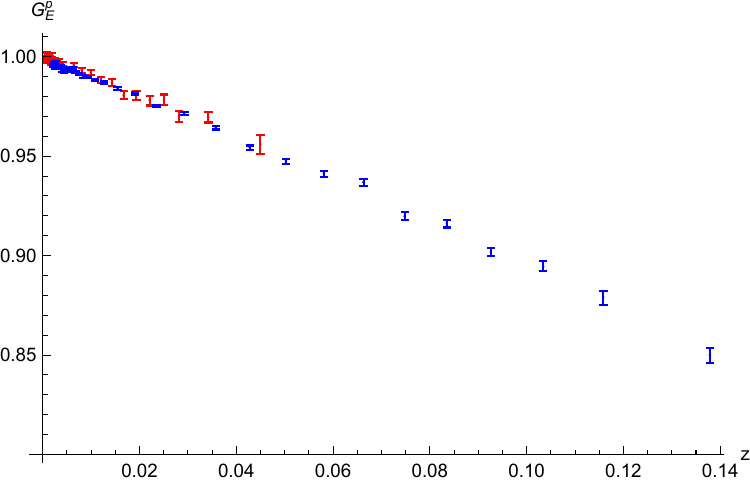}
\caption{\label{FigQ2Vsz} A comparison of $G_E^p$ from the PRad data set as a function of $Q^2$ (left) and $z$ (right). The 1.1 GeV (2.2 GeV) data set is in red (blue), the same color scheme used in \cite{Xiong:2019umf}. }
\end{center}
\end{figure}

Plotting the data as function of $z$ can be very instructive. For example, when only a slope can be constrained, the $z$-dependent data  will appear linear, while the $Q^2$-dependent data can appear to have curvature. See, e.g.,  figure 3 in \cite{Hill:2006ub} for a mesonic form factor and figure 2 in \cite{Epstein:2014zua} for a baryonic form factor. In figure \ref{FigQ2Vsz} we plot $G_E^p$ from the full PRad data set (described in the next section) as a function of $Q^2$ (left) and  $z$ (right). We can see a certain amount of curvature in the plot of $G_E^p$ as a function of $z$. We will explore this further in  section \ref{sec:ak}.

The default fit function used by PRad is 
\begin{equation}\label{Rational11}
f(Q^2)=nG_E^p(Q^2)=n\frac{1+p_1Q^2}{1+p_2Q^2}.
\end{equation}
In \cite{Xiong:2019umf} it is refereed to as ``Rational (1,1)". This function can be written as a sum of pole and a constant:
\begin{equation}
f(Q^2)=n\frac{1+p_1Q^2}{1+p_2Q^2}=n\frac{1-p_1t}{1-p_2t}=\dfrac{n(p_1-p_2)/p_2^2}{t-1/p_2}+\dfrac{n p_1}{p_2}.
\end{equation}
Provided that $1/p_2>4m_\pi^2$, this function's singularity lies above the two-pion threshold. In order to have the correct analytic structure, we must have  $1/p_2>4m_\pi^2$. We will check this requirement against PRad data in section \ref{sec:PRadDefault}. 

Assuming $1/p_2>4m_\pi^2$, the Rational (1,1) function can be expressed as a power series in $z$. The coefficients $a_k$ depend on its  imaginary part. Since this is a sum of pole and a constant, the imaginary part is a delta function. Using the expressions in \cite{Hill:2010yb} we find  
\begin{eqnarray}\label{rational11z}
a_0&=&f(0) \stackrel{\textnormal{Rational (1,1)}}{=}n \nonumber\\
a_{k \geq1}&=& {2\over \pi} \int_{t_{\rm cut}}^\infty {dt\over t-t_0} \sqrt{ t_{\rm cut} - t_0 \over t - t_{\rm cut}} 
\,{\rm Im}f(t)\sin[ k\theta(t) ]\stackrel{\textnormal{Rational (1,1)}}{=}\nonumber\\
&=&
\frac{2n(p_1-p_2)}{p_2}\sqrt{\frac{4m_\pi^2}{p_2^{-1}-4m_\pi^2}}\sin\left[k\cos^{-1}\left(1-8p_2m_\pi^2\right)\right].
\end{eqnarray}
We will compare these expressions to a $z$-expansion fit to the PRad data in section \ref{sec:ak}.

\section{PRad extractions of the proton charge radius} \label{sec:PRad}
Before improving on the $r_E^p$ extraction from the PRad data, we should be able to reproduce its published results.  We use the information in \cite{Xiong:2019umf} and its supplementary material. We use the PRad data release \cite{PRadData} from December 10, 2019. The ``raw" values of $G_E^p(Q^2)$ can be obtained from the ``1.1GeV\_table.txt" and ``2.2GeV\_table.txt" files, where they are  listed under ``f(Q2)".  The two files correspond to the 1.1-GeV and 2.2-GeV electron beams data of the PRad experiment.

We repeat many of the fits reported in \cite{Xiong:2019umf} and its supplementary material. We focus on the default PRad  Rational (1,1)  fit and  the fits involving the $z$ expansion.  We use the $\chi^2$ function
\begin{equation}\label{chi2PRdd}
\chi^2_\textnormal{ PRad}=\sum_{i=1}^N\dfrac{\left(G^{p,i}_{E,\mbox{\scriptsize exp.}}-G^{p,i}_{E,\mbox{\scriptsize theo.}}\right)^2}{\left(\delta G^{p,i}_{E,\mbox{\scriptsize exp.}}\right)^2},
\end{equation}
and minimize it for a given theoretical expression of $G_E^p$. We use only the statistical errors in $\delta G^{p,i}_{E,\textnormal{exp}}$. The proton charge radius is calculated via (\ref{rpDefined}). The uncertainty is found by the $\delta\chi^2=1$ range.  In reproducing the PRad fits we follow its  practice and include a normalization factor for the data as a multiplicative factor in $G^{p,i}_{E,\textnormal{ theo.}}$. This normalization factor is also determined from the fit. 

\subsection{Default PRad fit}\label{sec:PRadDefault}
The default expression for $G_E^p$ used by PRad is the Rational (1,1) function given in (\ref{Rational11}). The Rational (1,1) is fitted to both the 1.1 GeV and 2.2 GeV with different overall normalization factors called $n_1$ and $n_2$, but with the same $p_1$ and $p_2$. 

From our fit we find $n_1=1.0002\pm0.0002_\textnormal{statistical}$, $n_2=0.9983\pm 0.0002_\textnormal{statistical}$, and $r_E^p=0.831\pm0.007_\textnormal{statistical}$ fm. The reduced $\chi^2$ is 1.3. These are also the results in \cite{Xiong:2019umf}.  

As a further check, we find that the fit values  of $p_1$ and $p_2$ are $p_1=-0.0715$ GeV$^{-2}$ and $p_2=2.88$ GeV$^{-2}$. Up to the first three significant figures, these are the values reported in ``readme.pdf" in \cite{PRadData}. Including  the uncertainties on these parameters, we find $p_1=-0.07^{+0.56}_{-0.54}$ GeV$^{-2}$ and $p_2=2.88^{+0.61}_{-0.59}$ GeV$^{-2}$. Within the one standard deviation range we have $1/p_2>4m_\pi^2\approx 0.0784$ GeV$^2$. Thus the fit result is consistent with the analytic structure of $G_E^p$. 

\subsection{Other PRad fits}
In the supplementary material of \cite{Xiong:2019umf} the results of other fits to the PRad data are shown, but only in figures. Still, the approximate value of $r_E^p$ and its statistical uncertainty can be inferred from the figures. 

PRad performed fits to its  entire data set using a second order and third order polynomial in $z$. These correspond to truncating the series in (\ref {zseries}) at $z^2$ and $z^3$, respectively. For example, equation (2) of the supplementary material  is $f(Q^2)=nG_E^p(Q^2)=n(1+p_1z +p_2z^2)$. Using these expressions with different normalizations for the 1.1 GeV and 2.2 GeV data, our fit to the PRad data gives $r_E^p=0.830\pm0.008_\textnormal{statistical}$ fm for the second order polynomial in $z$ and  $r_E^p=0.825\pm0.015_\textnormal{statistical}$ fm for the third order polynomial in $z$.  These results agree with the values and statistical uncertainty in figure S15 of the supplementary material of \cite{Xiong:2019umf}. Notice also that the uncertainty is doubled when changing  from a second to a third order polynomial. We will address this problem below. 

PRad also performed  fits using Rational (1,1) to parts of the data set. These are listed in figure  S16(a) of the supplementary material of \cite{Xiong:2019umf}. Following PRad, we fitted the 1.1 GeV data, 2.2 GeV data, $Q^2<0.016$ GeV$^2$ data , and $Q^2>0.002$ GeV$^2$ data. We find $r_E^p=0.845^{+0.041}_{-0.039}$ fm for the 1.1 GeV data only, $r_E^p=0.829^{+0.008}_{-0.008}$  fm for the 2.2 GeV data only, $r_E^p=0.799^{+0.018}_{-0.017}$  fm for the $Q^2<0.016$ GeV$^2$ data, and $r_E^p=0.841^{+0.011}_{-0.011}$  fm for the $Q^2>0.002$ GeV$^2$ data. All uncertainties are statistical. These results agree with figure  S16(a).

Finally, PRad considered a fit of a second order polynomial in $z$ to the 2.2 GeV data only. Performing such a fit we find $r_E^p=0.829\pm0.009_\textnormal{statistical}$ fm. These results agree with figure  S16(b)  of the supplementary material of \cite{Xiong:2019umf}.

In conclusion,  we reproduced the values of $r_E^p$  reported by PRad from the PRad data. We now investigate if and how these results change when we use a model-independent extraction.  

\subsection{The need for a bound on the coefficients}\label{subsec:bound}
Truncating the $z$-expansion series, as was done in the PRad fits, might underestimate the uncertainty of $r_E^p$.  On the other hand, simply increasing the number of fitted parameters can overestimate the uncertainty. As shown in \cite{Hill:2010yb}, one needs to bound the coefficients. 

To illustrate that, we perform a fit to the PRad data of the form $f(Q^2)=nG_E^p(Q^2)=n(1+p_1z +p_2z^2+\cdots+ p_{k_\textnormal{max}} z^{k_\textnormal{max}})$. As in the PRad fits we use different normalization factors $n_1$ and $n_2$ for the 1.1 GeV and the 2.2 GeV data, but the same $p_k$ for both data sets. We consider two cases, no bound on $p_k$ and a bound $|p_k|<5$. We implement the bound as in \cite{Lee:2015jqa} by adding $\chi^2_\textnormal{Bound}=\sum_{k=0}^{k_\textnormal{max}} p^2_k/5^2$  to (\ref{chi2}). 

\begin{figure}
\begin{center}
\includegraphics[scale=0.6]{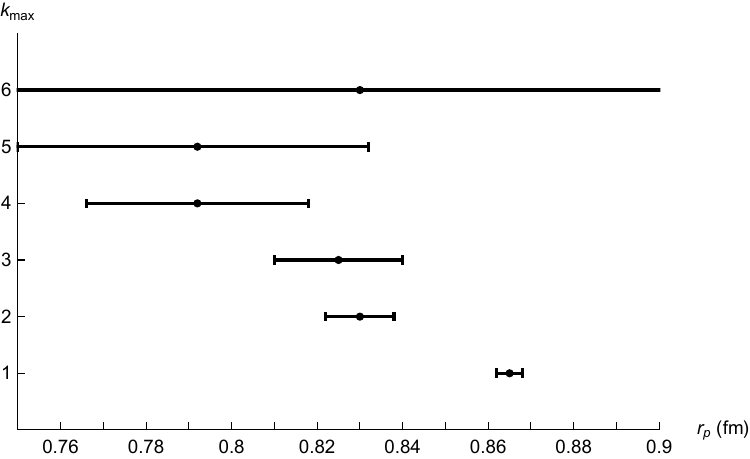}
\hspace{1cm}
\includegraphics[scale=0.6]{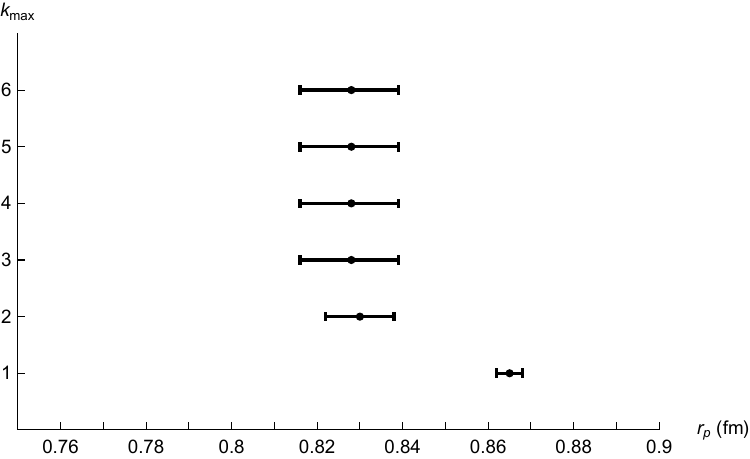}
\caption{\label{FigBounded} A comparison of the extracted $r_E^p$ as a function of the number of fitted $p_k$ parameters from $z$-expansion fits to the PRad data set. The range on the $x$-axis is the same range used in figure S15 of the supplementary material of \cite{Xiong:2019umf}.  Left: a fit without a bound on the coefficients $p_k$. Starting at $k_\textnormal{max}=5$ the uncertainty on $r_E^p$ exceeds the range $0.75-0.9$ fm. Right: a fit with a bound of 5 on the coefficients $p_k$.}
\end{center}
\end{figure}

The results of the two fits are shown in figure \ref{FigBounded} as a function of the number of fitted $p_k$ parameters.  As expected  \cite{Hill:2010yb}, the uncertainty on the extracted value of $r_E^p$ grows without bound  for the unbounded fit, while for the bounded fit it stabilizes on $r_E^p=0.828^{+0.011}_{-0.012}$ fm. 

\section{Model independent extraction of the proton charge radius} \label{sec:zfit}
Below we perform a model-independent $z$-expansion fit to PRad data, that includes a bound on the coefficients. We consider a fit to the whole PRad data set as well as the 1.1 GeV and 2.2 Gev data subsets. We also explore the effects of the bound on the coefficients, the $Q^2$ dependence of the extracted $r_E^p$, and the possible extraction of $a_k$ parameters beyond $a_1$.

\subsection{Model-independent $\bm z$-expansion $\bm{r_E^p}$ extraction from the entire PRad data}
We extract $r_E^p$ from the PRad data by using the $\chi^2$ function
\begin{equation}\label{chi2}
\chi^2_z=\sum_{i=1}^N\dfrac{\left(\eta_i G^{p,i}_{E,\textnormal{ exp.}}-G^{p,i}_{E,\textnormal{ theo.}}\right)^2}{\left(\delta G^{p,i}_{E,\textnormal{ exp.}}\right)^2}.
\end{equation}
As before, $G^{p,i}_{E,\textnormal{ exp.}}$ are the values of $G_E^p$ reported by PRad and $\delta G^{p,i}_{E,\textnormal{ exp.}}$ its statistical errors. $G^{p,i}_{E,\textnormal{  theo.}}$ is given in (\ref{zseries}) where the series is truncated at $k_\textnormal{max}$.  The normalization factor are  $\eta_i=\eta_1$ if $i$ is part  of the 1.1 GeV data set, and $\eta_i=\eta_2$ if $i$ is part  of the 2.2 GeV data set. Thus we allow for a normalization factor for each data set, but unlike PRad fits, we do not include it in $G^{p,i}_{E,\textnormal{  theo.}}$. Therefore this $\chi^2$ function differs from the one used in section \ref{subsec:bound}. Since we include a normalization factor, we fix $G_E^p(0)=1$ which implies $a_0=1$ in the fits. 

In order to bound the coefficients we add to $\chi^2_z$, as in \cite{Lee:2015jqa}, $\chi^2_\textnormal{Bound}$ defined as 
\begin{equation}\label{chi2Bound}
\chi^2_\textnormal{Bound}=\sum_{k=0}^{k_\textnormal{max}} \frac{a^2_k}{B^2},
\end{equation}
where $B$ is a pure number. Our default value is $B=5$, but we check our results also for $B=10$. This choice of bounds was discussed in detail and implemented in the literate, see \cite{Hill:2010yb, Epstein:2014zua, Lee:2015jqa} for the nucleon electromagnetic form factors and \cite{Bhattacharya:2011ah, Bhattacharya:2015mpa, Meyer:2016oeg} for the nucleon axial form factor.

Fitting the entire PRad data with $B=5$, we find that the extracted proton charge radius is $r_E^p=0.828^{+0.011}_{-0.012}$ fm. Changing the bound to $B=10$ we find $r_E^p=0.827^{+0.013}_{-0.014}$ fm.  These are almost identical one-standard deviation ranges. Compared to the default PRad fit of Rational (1,1), $r_E^{p,\textnormal{rational}}=0.831\pm0.007$ fm, the central values are almost the same, but the uncertainty is more than 50\% larger for the $z$-expansion fit.  The extracted $r_E^p$ stabilizes for $k_\textnormal{max}=3$. It does not change as we increase $k_\textnormal{max}$ above 3. We  checked this with fits up to $k_\textnormal{max}=10$. 

As another check, we consider fits without adding  $\chi^2_\textnormal{Bound}$ and by using explicit bounds  of $|a_k|\leq 5$ and $|a_k|\leq 10$ as in \cite{Hill:2010yb, Epstein:2014zua}. We find $r_E^p=0.824^{+0.015}_{-0.012}$ fm for $|a_k|\leq 5$ and $r_E^p=0.824^{+0.015}_{-0.015}$ fm for $|a_k|\leq 10$.  These are very close to the results above that used $\chi^2_\textnormal{Bound}$.

\subsection{Model-independent $\bm z$-expansion $\bm{r_E^p}$ extraction from parts of the PRad data}
The  Rational (1,1)  fits to the 1.1 GeV and 2.2 GeV parts of the PRad data give values of $r_E^p$ that are within the one-standard deviation range of each other, but the uncertainty on the former is five times as large. It is instructive to see what are the results for $z$-expansion fit. We use the same fit $\chi^2$ function, namely the sum of $\chi^2_z$ and $\chi^2_\textnormal{Bound}$. 

Using only the 1.1 GeV data set we find $r_E^p=0.847^{+0.035}_{-0.037}$ fm for $B=5$ and $r_E^p=0.846^{+0.039}_{-0.041}$ fm for $B=10$. These  are almost identical to  the Rational (1,1)  fit result of  $r_E^{p,\textnormal{rational}}=0.845^{+0.041}_{-0.039}$ fm.  Using only the 2.2 GeV data set we find $r_E^p=0.826^{+0.012}_{-0.013}$ fm for $B=5$ and $r_E^p=0.823^{+0.015}_{-0.015}$ fm for $B=10$. These are consistent with the Rational (1,1)  fit result of  $r_E^{p,\textnormal{rational}}=0.829^{+0.008}_{-0.008}$ fm, but the uncertainty is more than 50\% larger for the $z$-expansion fit. 

Another question we study is the effect of a cut on $Q^2$. We consider this question for the 1.1 GeV data alone,  the 2.2 GeV data alone, and the combined 1.1 GeV and 2.2 GeV data. We perform fits to $r_E^p$ for data with $Q^2$ smaller than  $Q^2_\textnormal{cut}$. The lowest $Q^2_\textnormal{cut}$ is determined by the requirement that the slope of the form factor is positive. If $Q^2$ is too small, we do not have enough data for a meaningful extraction of $r_E^p$.  Thus $Q^2_\textnormal{cut}\in[0.0007, 0.0155]$ GeV$^2$ for the 1.1 GeV data set, $Q^2_\textnormal{cut}\in[0.001, 0.059]$ GeV$^2$ for the 2.2 GeV data set, and $Q^2_\textnormal{cut}\in[0.0007, 0.059]$ GeV$^2$ for the combined 1.1 GeV and 2.2 GeV data sets. All fits are performed with $B=5$.

\begin{figure}[h]
\begin{center}
\begin{center}
\includegraphics[scale=0.6]{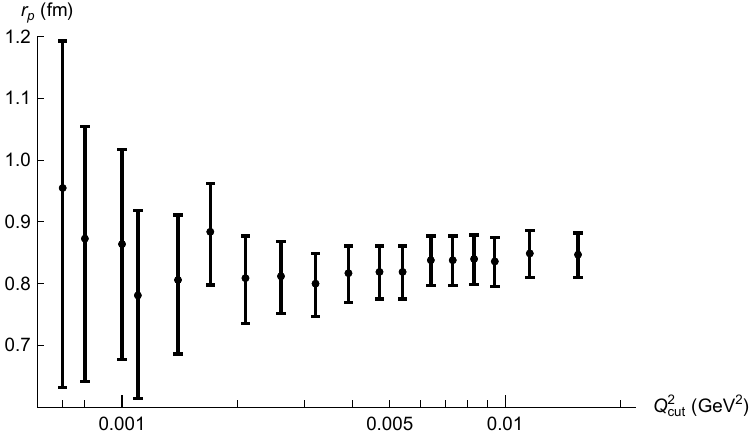}
\hspace{1cm}
\includegraphics[scale=0.6]{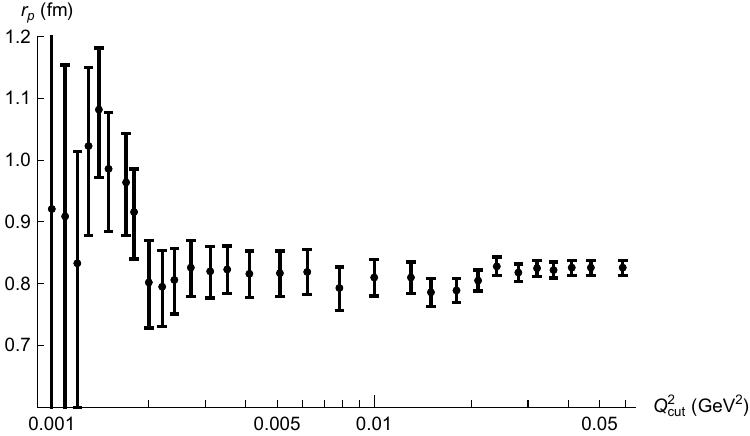}
\end{center}
\begin{center}
\includegraphics[scale=0.6]{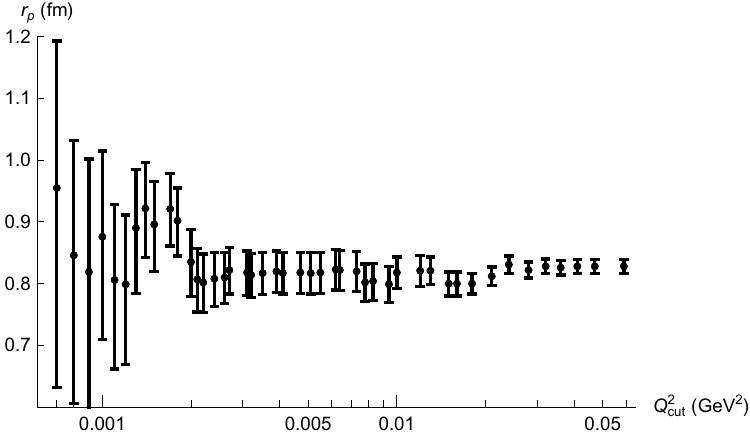}
\end{center}
\caption{\label{FigQ2Cut} Extracted $r_E^p$ as a function of $Q^2_\textnormal{cut}$.  Top left: 1.1 GeV data set. Top right: 2.2 GeV data set. Bottom center: 1.1 and 2.2 GeV data sets.}
\end{center}
\end{figure}

The results of the extractions are shown in figure \ref{FigQ2Cut}. In all three plots we see a convergence to a value as $Q^2_\textnormal{cut}$ is increased. In the 2.2 GeV data set plot and the combined 1.1 GeV and 2.2 GeV data sets we see also a ``peak" at about $Q^2_\textnormal{cut}=0.0014$ GeV$^2$. But overall the extraction is independent of the cut on $Q^2$, for large enough $Q^2_\textnormal{cut}$.

\subsection{Model-independent $z$-expansion fit to the entire PRad data}\label{sec:ak}

The charge radius is only a one-parameter characterization of the data. We can try and extract more coefficients in  (\ref{zseries}). To do that we use (\ref{chi2}) and add to it a modified version of (\ref{chi2Bound}) where we omit the term $a_i^2/B^2$ in the sum in (\ref{chi2Bound}) when constraining the $a_i$ coefficient.  

We perform such a fit to the full PRad data set (both 1.1 and 2.2 GeV) using $B=5$. The fit stabilizes quickly for $k_\textnormal{max}=i+3$. We find that $a_1=-0.921\pm0.026$, $a_2=-1.2\pm0.6$, and $a_3=2.2\pm5.7$.  Using $B=10$ gives very similar results. This implies that beyond a slope ($a_1$), only a curvature ($a_2$) can be obtained from the PRad data\footnote{We can compare these values to the values predicted by the Rational (1,1) fit from section \ref {sec:PRadDefault}. Using $p_1=-0.07^{+0.56}_{-0.54}$ GeV$^{-2}$ and $p_2=2.88^{+0.61}_{-0.59}$ GeV$^{-2}$, equation (\ref{rational11z}) gives $a_1^\textnormal{ Rational (1,1)}=-0.93\pm0.26$ and $a_2^\textnormal{ Rational (1,1)}=-1.02\pm0.19$. These agree with the values we obtained from the $z$-expansion fit.}.

\begin{figure}[h]
\begin{center}
\includegraphics[height=11em]{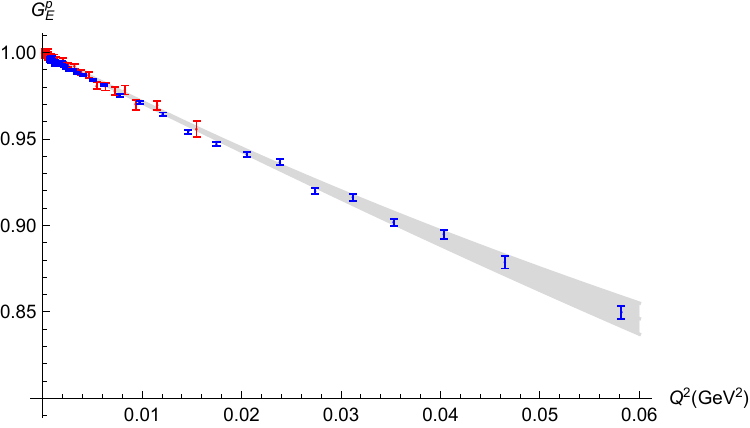}
\hspace{1cm}
\includegraphics[height=11em]{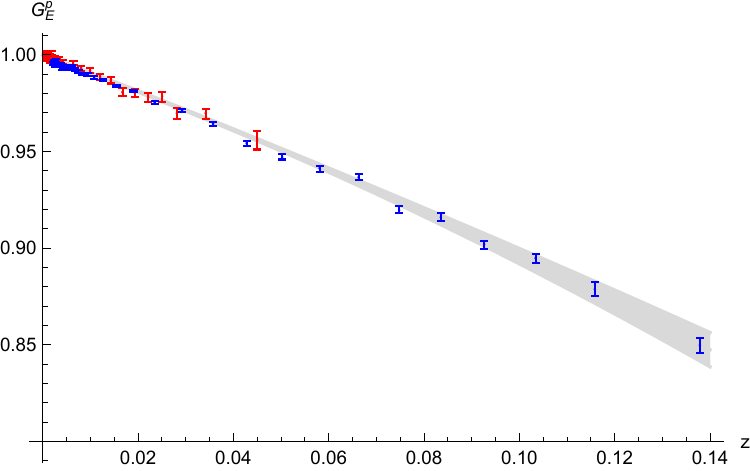}
\caption{ \label{FitQ2Vsz} A model-independent $z$-expansion fit to $G_E^p$ from the entire PRad data set (gray band) together with values of $G_E^p$ from the PRad data set as a function of $Q^2$ (left) and $z$ (right). The 1.1 GeV (2.2 GeV) data set is in red (blue), the same color scheme used in \cite{Xiong:2019umf}.}
\end{center}
\end{figure}

To compare these results graphically to the PRad data, we perform a fit with $B=5$ to the full PRad data set without bounding $a_1$ and $a_2$, i.e.,  we omit the terms $a_1^2/B^2$ and $a_2^2/B^2$ in the sum in (\ref{chi2Bound}). The fit stabilizes quickly with increasing $k_\textnormal{max}$. We find the values above for $a_1$ and $a_2$ and a covariance of $-0.0137$ between them. The variance of the data normalization factors $\eta_{1,2}$ is negligible as well as their covariance with $a_1$ or  $a_2$.  The resulting fit and uncertainty \cite{Pruneau:2017ypa} is shown in figure \ref{FitQ2Vsz} together with the PRad data from figure \ref{FigQ2Vsz}.

\section{Conclusions} \label{sec:conclusions}
The proton radius puzzle has motivated new theoretical and experimental work. Among them is PRad, a new electron-proton scattering experiment at Jefferson Lab.  PRad reached the lowest $Q^2$ in $e-p$ scattering: $2.1\times10^{-4}$ GeV$^2$,  an order of magnitude lower than previously achieved at A1 Mainz \cite{Bernauer:2010wm,Bernauer:2013tpr}. The small $Q^2$ should allow to reduce extrapolation errors in extracting the proton charge radius.

PRad has extracted a radius of $0.831\pm 0.007_\textnormal{statistical}\pm 0.012_\textnormal{systematic}$ fm by using a Rational (1,1) fit function for $G_E^p$. Instead of relying on a specific model for $G_E^p$, one can use a model-independent approach via the $z$ expansion. In this paper we have examined how the statistical error reported by PRad changes when using such a model-independent approach. 

In section \ref{sec:PRad} we repeated many of the fits performed by PRad to its data. These include its default Rational (1,1) fit to the entire PRad data, a second and third order polynomial in $z$  fit to the entire PRad data, Rational (1,1) fit to parts of its data set, and a second  order polynomial in $z$  fit to its 2.2 GeV data. These agree with \cite{Xiong:2019umf}, its supplementary material, and information from the PRad data release \cite{PRadData} from December 10, 2019. 

We also compared the extractions of $r_E^p$ when higher polynomials in $z$ are considered, with and without bounding the coefficients of the polynomial in $z$. The results appear in figure \ref{FigBounded}. As expected \cite{Hill:2010yb}, we find that the extracted proton charge radius grows without bound for the unbounded fit, while for the bounded fit it stabilizes very quickly to a value independent of the degree of the $z$ polynomial.   

In section \ref{sec:zfit} we performed a model-independent $z$-expansion fit to PRad data.  The bounding of the coefficients is implemented by adding a term to $\chi^2$ \cite{Lee:2015jqa}, see (\ref{chi2Bound}). From a fit to the entire PRad data set we find $r_E^p=0.828^{+0.011}_{-0.012}$ fm. Compared to the default PRad fit, $r_E^{p,\textnormal{rational}}=0.831\pm0.007$ fm, the central values are almost the same, but the uncertainty is more than 50\% larger for the $z$-expansion fit. This implies that PRad's default fit underestimates the statistical error by using the Rational (1,1) function. 

We also performed a model-independent $z$-expansion fit to parts of the PRad data. We fitted the 1.1 GeV and 2.2 GeV parts of the PRad data separately. For the 1.1 GeV data (that contains the smaller $Q^2$ data) we find that the model independent extraction is almost identical to the  Rational (1,1) fit. The error bar of this extraction is too large to distinguish between the two values of the proton charge radius. For the 2.2 GeV data the model independent extraction uncertainty is 50\% larger than the Rational (1,1) fit.  We considered also the effects of a $Q^2$  cut on the data, $Q^2 <Q^2_\textnormal{cut}$. The results are shown in figure \ref{FigQ2Cut} for the 1.1 GeV data, the 2.2 GeV data, and the entire PRad data.  Overall the extraction is independent of the cut on $Q^2$, for a large enough $Q^2_\textnormal{cut}$. 

Going beyond $r_E^p$, we fitted more parameters in the $z$ expansion to the PRad data. The results are described in section \ref{sec:ak} and figure \ref{FitQ2Vsz}. We find that beyond  the slope, equivalent to $r_E^p$,  only a curvature can be obtained from the PRad data.  

Before concluding, let us briefly review recent papers that also analyzed the PRad data. In \cite{Horbatsch:2019wdn} PRad data was analyzed to investigate their consistency with $r_E^p$ from muonic hydrogen and theoretical predictions for the coefficients of $Q^4$ and $Q^6$ terms in the $Q^2$ expansion of $G_E^p$. Using a rational function to incorporate these inputs, the author of \cite{Horbatsch:2019wdn} found very good agreement with the PRad data. In \cite{Alarcon:2020kcz} a fit using the DI$\chi$EFT model to the PRad and A1 Mainz data \cite{Bernauer:2010wm,Bernauer:2013tpr} was performed. The authors of  \cite{Alarcon:2020kcz} found the same value of $r_E^p$ within uncertainties as their fit to A1 Mainz data alone. Finally, very recently \cite{Borah:2020gte} appeared that compared fits using the $z$-expansion to non-PRad scattering data and PRad data. The authors of  \cite{Borah:2020gte} remark that their $z$-expansion fit to PRad data, taking the PRad errors at face value, results in a significantly larger uncertainty for $r_E^p$ compared to the Rational (1,1) PRad fit. 

In summary, using model-independent methods we find that the statistical uncertainty on the proton charge radius from the PRad data is more than 50\% larger than the one quoted by PRad in \cite{Xiong:2019umf}. The systematic error is obtained by a much more involved process that is described  in the supplementary material of the PRad paper \cite{Xiong:2019umf}. It is likely that the systematic error will also increase when using model-independent methods. It is needed for a full model-independent extraction of the proton charge radius from the PRad data.

\vskip 0.2in
\noindent
{\bf Acknowledgements}
\vskip 0.1in
\noindent
We thank Haiyan Gao, Claude Pruneau, and Weizhi Xiong for useful discussions. This work was supported by the U.S. Department of Energy grant DE-SC0007983 and by a Career Development Chair award from Wayne State University.

\end{document}